\documentstyle[aps,epsf,epsfig]{revtex}
\begin{document}
\input epsf
\title{Classical Fields and the Quantum concept}
\author{Manoelito M de Souza}
\address{Universidade Federal do Esp\'{\i}rito Santo - Departamento de
F\'{\i}sica\\29065.900 -Vit\'oria-ES-Brasil}
\date{\today}
\maketitle
\begin{abstract}
We do a  critical review of the Faraday-Maxwell concept of classical field and of its quantization process. With the hindsight knowledge  of the essentially quantum character of the interactions, we use a naive classical model of field, based on exchange of classical massless particles, for a comparative and qualitative analysis of the physical content of the Coulomb's and Gauss's laws. It enlightens the physical meaning of a field singularity and of a static field. One can understand the problems on quantizing a classical field but not the hope of quantizing the gravitational field right from General Relativity.
\end{abstract}
\begin{center}
\end{center}

\begin{section}
{INTRODUCTION}
\end{section}
\noindent  The concept of interacting field introduced by Faraday-Maxwell 
is on the basis of Field Theory and of Quantum Mechanics. It is necessary 
for a relativistic description of the wave-particle duality of the quantum theories. The idea of a classical interaction as a wave, continuous and distributed over the whole space, is put against the modern idea of a 
quantum interaction, discrete and localized in ``corpuscles" or interaction quanta. The passage from the first to the second idea requires a so called quantization process, and this passage, it is well known, in the best of 
the cases (QED) has deep problems, and in the worst case (QG) it has proved to be not viable. In the search of solutions to this problem of quantizing the gravitational field the observed tendency is the one of replacing 
complex models and formalisms for others of increasing complexity. On the other hand, the Classical Electrodynamics, the best known and 
``well succeeded" paradigm of all subsequent field theory, has old and well known problems of inconsistency with the description of fields in the neighborhood of their sources. In a recent work (hep-th/9610028) we consider these problems and we show that taking the correct zero distance limit to the charge reveals unequivocal clues of quantum features in Classical Electrodynamics: the flux of field from a charge is discontinuous in time. Out of this limit this discontinuity is masqueraded by the field spacetime-average character. The problems of CED in its zero distance limit, which are erroneously attributed to the working hypothesis of a pointlike electron, are rather unequivocal signs that our concepts of fields are unappropriate for describing interactions. The idea of a continuous classical field is misleading; it is valid only for large distances and large number of photons.\\
 \noindent We want to make here a critical review of the Faraday-Maxwell concept  of classical field under the perspective of modern physics that understands it as being of a fundamentally quantum nature. It is well known that the pioneers of classical field theory worked with a  model for the electromagnetic phenomena based on an analogy with the fluid mechanics. The electromagnetic effects would be propagated through an all-pervading fluid, the ether. This, at the time, new vision of field-mediated interactions between distant charges was an advance with respect to the Newtonian 
concept of action at a distance. The fluid analogy implies on a field distributed 
all over the space around its source, like it happens with the sound waves, for example. This image was certainly reinforced by what at that time 
seemed to be an apparently definitive victory of the concept of light as a wave phenomena against the Newtonian model of light as a stream of corpuscles. The phenomena of interference, diffraction and polarization of light had been decisive for this conviction. Only much later the first 
clues of a discreteness, like the photoelectric and the Compton effects, would be discovered. But even Quantum Mechanics that was created from the necessity of explaining this new kind of effects received also the influence of this fluid-mechanics concept of field: the wave function is a space-distributed field representing a fluid of amplitude of probabilities. In this qualitative analysis we want to oppose this historical vision of classical fields as some waves, continuous and distributed all over the space, against the vision of interactions mediated by localized point-like objects, their quanta, discretely emitted, propagated and absorbed. In the modern perspective all the four fundamental interactions of nature are mediated by their respective  quanta. We will try to obtain a qualitative view on how a classical field theory could be formulated  if the 
interaction were seen as mediated by massless point-like objects 
propagating on straight-line trajectories between their emitters and 
their absorbers.\\
 
\begin{section}
{The Gauss's law and the singularity problem}
\end{section}
The Coulomb's law (or in the case of gravitation, the Newton's law) describes the {\bf effectively observed} interaction between two static electric charges in terms of forces acting on the charges along the straight-line defined by the charges positions and with a magnitude that is directly proportional to the product of the two charges and to the inverse of the square of the distance between them. The observed spherical symmetry 
around a point charge is the assurance that nothing changes in the above description if the second point charge is moved to any other point of a spherical surface centered at the first charge. The forces acting on the charges are  what is actually observed and they require the presence of
 the two charges and they are observed only at the charges sites.\\ The concept of a field existing everywhere around a single charge, regardless
 the presence of any other charge is an extrapolation of what is 
effectively observed. There is, therefore, a very deep distinction between the physical content of the  Coulomb's and of the Gauss's laws.
This last one describes the {\it inferred} electric field as existing around a single charge, independent of the presence of the other charge.
The electric field, as it is well known, is extracted from the Gauss's law
through the integration of its flux across a surface, having the
 appropriate symmetry, {\it enclosing} the charge,
\begin{equation}
\label{Gauss}
{\vec E}(x)={\hat e} \frac{\int^{x}_{V}\rho dv}{\int_{\partial V}dS},
\end{equation}
where ${\hat e}$ is the unit vector normal to the closed surface 
$\partial V.\;$
(\ref{Gauss}) puts in evidence the effective or average character of
the Faraday-Maxwell's concept of field; it gives also a hint on the meaning and origin of the field singularity. If the electric field can be visualized 
in terms of exchanged photons, then according to (\ref{Gauss}), the 
frequency or the number of these exchanged photons must be proportional to the enclosed net charge. And if we take ${\vec E}$, as suggested by the Gauss's law, as a
measure of the flux of photons emitted/absorbed by a point charge, we can schematically write, $E\sim \frac{n}{4\pi r^{2}},$ where n is the number 
of photon per unit of time crossing an spherical surface of radius r and centered on the charge. Then, the divergence of E in the limit when $r\rightarrow0$ does not represent a physical fact like an increasing 
number of photons, but just an increasing average number of photons per 
unit area, as the number of photons remains constant but the area tends to zero. So, a field singularity would have no physical meaning, because it would just be a consequence of this average nature of the Faraday-Maxwell's field, and 
contrary to what is usually thought, it would not be a consequence  of the 
electron point-like nature. 

Taken the exchange of ``quanta" (here in the sense of discrete and 
localized chunks of energy and momenta, like in a classical particle) 
as a model for a classical interaction one must conclude that the Faraday-Maxwell concept of field, which lies behind the justification 
of the use of the Gauss's law, must be seen as the smearing of the effect 
of the ``quantum"-exchange on the whole space around a charge, and 
during  a time interval larger than the time (period) between two 
consecutive   exchange of a ``quantum". Under this perspective, the Faraday-Maxwell concept of a classical field corresponds in fact to an average in space and in time of the actual quantum interaction. It 
replaces something discrete in space and in time and localized on the straight-line between the two interacting charges by something 
isotropically spread in space 
around each charge and in time. The quantization process is a tentative 
of reversing this operation. Does it make sense, in this new perspective? 
Is it the more appropriate approach? The answer to both questions, in this context, seems to be no. It seems to be more appropriate to reformulate the description of classical physics in terms of discrete interactions before trying to quantize it. 
The two concepts (classical and quantum) of interaction are associated to domains with distinct topologies: the light-cone and the straight-line, respectively. The quantization process does not account for this 
difference. The classical field, as a massless wave, propagates on the light-cone, which is not a manifold because of the singularity on its vertex. The quantum interaction, on the contrary, is defined on a 
light-cone generator, a straight-line, and therefore, has no singularity.\\ 

\begin{section}
{ Quantum Gravity and General Relativity}
\end{section}

The problems with the description of fields in a close vicinity of their sources seem to be that we are taking the fields by their averages. This seems to be true for the electromagnetic field  and it may also be true for the other known classical interaction, the gravitation.
For the gravitational field there is a further complication given by the General Theory of Relativity which replaces the description of a gravitational force by the picture of a curved (Riemannian) spacetime. In the context of discrete interactions, this geometrization is an added averaging process as it changes a polygonal trajectory of an interacting test mass smoothing it into a geodesic. With discrete interactions, the events of emission and absorption of a gravitational quantum by a mass form the vertices of a polygonal trajectory; they  are connected by  straight-line segments that correspond to the,  in between, mass's free motion. 
This geometrization not only hides the interaction 
discreteness as it also makes more difficult any tentative of retrieving it by quantization, since it mixes the geometry of the background Minkowski spacetime (its metric)  with the actual physical effect (the exchange of quanta) incorporating them into the metric tensor of a Riemannian manifold. To quantize this metric tensor would be, therefore, tantamount to a discretization 
not only of the gravitational interaction but also of the spacetime. The physicists who see this picture of a curved spacetime not as an
 approximation but as a fundamental aspect of nature does not, of course, agree with this; but in this context of discrete interactions any whole-metric quantization does not make sense.\\ 
 The Einstein field equations, like the Maxwell's equations, deal with spacetime-average fields.
The General Theory of Relativity has a solid basis of experimental confirmation, but like Classical Electrodynamics, only for situations where the classical approximation of the field as an spacetime average is a good description: far from their sources and involving a large number of quanta.
In this perspective of discrete classical interactions, one can understand the nature of the problems that show up in the quantization of the classical fields but it seems then that there is no justification for a hope that the gravitation field may be quantized starting from the General Theory of Relativity.

This is more than just an academic discussion. If such observations are correct and if the above considerations about the physical meaning of the classical field singularities are valid also for 
the gravitational field of Einstein, described by a metric tensor, one must worry about the enormous intellectual efforts that are being devoted  to a detailed comprehension of  black-hole physics. It is opportune to 
remember that all known indirect evidences of black-holes are just indications of  possibly very intense gravitational fields produced by 
very compact objects but not necessarily black-holes.\\

\begin{section}
{The Coulomb's law and the meaning of static field}
\end{section}
The Gauss's law, in a picture of continuous interaction,  has a natural explanation for the 
dependence of the interaction with the inverse of the squared 
distance as a consequence of the 3-dimensionality of the space. The 
origin of this $\frac{1}{r^{2}}$-dependence must be entirely distinct
 if the interactions are seen as the result of exchange of particles.

This is a nagging problem in this picture of a classical 
field in terms of discrete interactions, which is also related, in the context of an actual quantum field theory, to the conceptual meaning of a static field. The $\frac{1}{r^{2}}$-dependence of the static force between two point charges (or masses) is an experimental fact as stated in the Coulomb's law (or Newton's law). In the context of the present analysis the question now is how to understand this $\frac{1}{r^{2}}$-dependence as well as the meaning
 of a static field in terms of a discrete exchange of particles. In QFT theory one deals with quantum fluctuation and virtual-particle exchange. Here, in a classical context, there is no virtual particle and no quantum 
fluctuation. All particles are real with positive and definite energy
 and mass (which may be null). Strict conservation laws for energy and (angular and linear) momentum must be always observed.

But before proceeding further on this, we must remind some well-known experimental facts:
\begin{enumerate}
\item Only an accelerated charge can radiate. A non-accelerated charge 
never radiates.\\
On the other hand, as a consequence of the energy-momentum conservation,
 the act of emitting or absorbing radiation by a charge necessarily 
results on 
its acceleration. So, for a charge, we can put it this way:
$${\hbox{RADIATION}}\Leftrightarrow {\hbox{ACCELERATION}}$$
\item A charge under an external periodic stimulation (force) emits 
radiation with the same frequency of the external stimulus. Or:
$${\hbox{FREQUENCY OF THE EMITTED RADIATION}}={\hbox{FREQUENCY OF THE EXTERNAL FORCE}}$$
\end{enumerate}
Accepting the above empirical fact 1 as being also valid in a fundamental level
 has some immediate logical implications:
\begin{itemize}
\item An isolated charge will never radiate, will never emit a single 
photon, will never be accelerated or will never accelerate itself (by emitting/absorbing a photon, as it is isolated).
\item So, it does not make sense talking of the electric field of a single isolated charge (taken as in the elementary pedagogical hypothesis of the only  existing charge in the world and not under any external force).
\item This is contrary to the ideas of quantum fluctuations (in QFT) and 
of self-interactions.
\item The fundamental (in the sense of irreducible) electromagnetic interaction does not correspond to a 3-legs Feynmann diagram (which does 
not obey the conservation laws, anyway) but to a connected set of two of them, accounting for the fact that an electron must be stimulated (accelerated by the absorption of a photon) to irradiate (emit a photon). See the figure 1. 
\end{itemize}
 For both pictures of interaction, mediated by a continuous and 
distributed field or discretely produced by the exchange of particles, 
some external forces must be provided to assure that the 
charges remain in a static equilibrium. The first immediate distinction 
is that in the continuous case the force must act continuously at both charges while in the particle-exchange picture the forces act only in the brief instant of emission and absorption of a quantum, as a kind of elastic reaction force.  See figure 2, where we are not neglecting a possible time delay, $\delta t$, between the absorption and the consequent re-emission of a photon.\\Each charge then emits a photon after 
being stimulated (accelerated) by the absorption of a photon emitted by 
the other charge.
There is a much diffused false premise  that the wave concept of field, 
in classical field theories, or of the emission followed by the 
reabsorption of virtual quanta, in quantum field theory, are necessary for explaining the interaction between two separated charges, otherwise, it is argued, how could the charges know the presence of each other? But this is indeed a not well posed question. What we really know from our best theories (and confirmed by  experiment) is that interactions and the emission and absorption process are closely interdependent concepts: an
 electric charge does not radiate unless it is under the action of an external force (interaction) and, according to our modern understanding, 
any change in the state of motion of the charges is a consequence of the exchange (emission or absorption) of quanta. It is not a question of what comes first, the emission or the interaction: they are just two aspects of
 a same thing. A complete understanding of this may be a subject of scientific investigation in the future; today this question still belongs to 
the domain of plain philosophy.  All we can say now, as physicists, is that 
a charge is accelerated with the emission and absorption of radiation and that
 it radiates only when accelerated. In quantum mechanics the condition that
 a charge be accelerated without emitting electromagnetic radiation leads 
to stationary states or the quantization of its energy.

Let us consider an imaginary Coulomb's experiment for measuring the 
repulsive force between two electrons.
Let r be the separation between the two point electric charges, fixed in 
their positions, each one, by a dynamometer on which one can read the value of the applied force on each electron. There is a force acting on a charge just in the instants when it receives a photon (emitted by the other electron) and emits also (as a reaction) a photon in the direction of the other electron. So, each dynamometer indicates a discrete, instantaneous force, discontinuous but periodical, with a period T,$$T=\frac{2r}{c},$$ where c is the speed of light. According to our listed experimental fact number 2, each electron, stimulated by this periodical force, emits photons with this same frequency, $f=\frac{c}{2r}.$ This seems to introduce a non-locality in the electrostatic interaction because the emitted photon depends on the distance between the interacting charges. But this is just apparent because this dependence on r comes from the photon's (two-ways) travelling time between the charges.
 Using the quantum information (the de Broglie's relation:$E=hf=h/T$), we have for the change in the momentum  
of an electron, $\Delta p$, during a time interval $\Delta t,$
\begin{equation}
\Delta p=(\frac{2hf}{c})(f\Delta t)=\frac{2hf^{2}\Delta t}{c},
\end{equation}
or
\begin{equation}
\label{fl}
 F=\frac{\Delta p}{\Delta t}=\frac{hc}{2r^{2}}.
\end{equation}
Then, with $\alpha=\frac{e^{2}}{4\pi\hbar c}$, in rationalized Gaussian units, we have 
\begin{equation}
\label{cl}
 F=\frac{1}{4\alpha}\frac{e^{2}}{r^{2}},
\end{equation}
or 
\begin{equation}
\label{nl}
 F=\frac{1}{4\alpha_{g}}\frac{Gm^{2}}{r^{2}}
\end{equation} with $\alpha_{g}=\frac{Gm^{2}}{4\pi\hbar c}.$
(\ref{cl}) and (\ref{nl}) are, respectively, the Coulomb's law and the Newton's gravitation law, up to a multiplicative constant. It is amazing that all of their qualitative aspects can be so easily obtained from such a simple and naive model of interaction. The two laws describe interactions between two static sources, but while the Coulomb's law gives an exact description, the Newton's law is just a weak field approximation. This difference may be explained by the mixing of the Minkowski metric with the graviton-exchange effects in the Einstein's field and by its space average nature. The attractive or repulsive character of the electrostatic  force requires the use of angular momentum conservation in which an essential role (like 
in QFT) belongs to the photon's spin. For simplicity, as this is not the point on the present focus, we will neglect the particles' spins. They 
will be treated as scalar objects. Only linear momentum conservation is involved and this only leads to repulsive force. The point here is to understand the  $\frac{1}{r^{2}}$-dependence of the force.\\
The spin of the exchanged quantum makes the differences among  distinct 
kind of interactions (scalar, vectorial and tensorial). For the validity of (\ref{cl}), $\Delta t>>T>>\delta t,$ must be satisfied.  As the distance between the two interacting charges in any Coulomb's experiment is of the order of centimeters, the lapse of time (T) between two consecutive emissions is about $10^{-10}$ s. So, $\Delta t>>10^{-10}$ s. This is in the reach of an experimental detection, and so, it would be possible, at least in principle, to observe discrete interactions with a carefully done Coulomb's experiment.

In summary, according to this view of interactions as exchanges of quanta,  both statements, the Gauss's law and the Coulomb's law, correspond to smoothing approximative averages hiding the process natural discreteness:
 the Gauss's law is an average in space and time while the Coulomb's law 
is an average in time. But they are not equivalent as they produce distinct consequences. The space average causes a topological change as it replaces the action of a single quantum, which propagates along a light-cone 
generator by a continuous field or wave propagating on the light-cone. It replaces, therefore, the simple topology of a straight-line (the light-cone generator, domain of the quantum) by the topology of the light-cone (domain of the wave). The light-cone is not a manifold because of its singularity on its vertex which is a reflection of the wave singularity.
\newpage
\vglue23cm
\begin{figure}

\epsfbox[0 0 30 50]{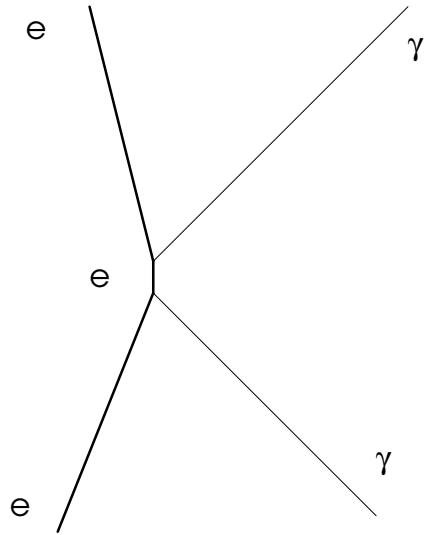}
\vglue-17cm
\caption[Fig. 1.]{Classical picture of the fundamental quantum process: an electron must be stimulated (accelerated) by the absorption of a photon in order to emit a photon.}
\end{figure}
%
\vglue25cm
\begin{figure}

\epsfbox[0 0 30 50]{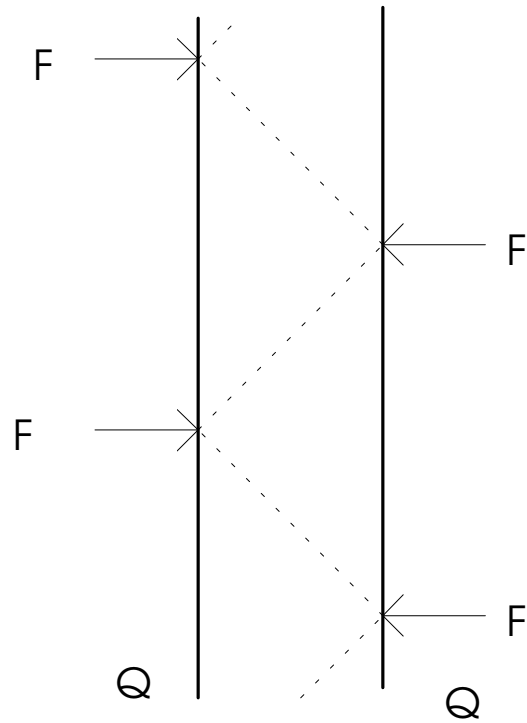}
\vglue-15cm
\caption[Fig. 2.]{The Coulomb's experiment or the static field seen as the exchange of real photons between two fixed point charges.}
\end{figure}

\end{document}